\documentclass[twocolumn,conference]{IEEEtran}
\usepackage[T1]{fontenc}
\usepackage[latin9]{inputenc}
\usepackage{float}
\usepackage{amsmath}
\usepackage{amsthm}
\usepackage{amssymb}
\usepackage{graphicx}
\makeatletter

\newcommand{\noun}[1]{\textsc{#1}}
\providecommand{\tabularnewline}{\\}

\theoremstyle{plain}
\newtheorem{thm}{\protect\theoremname}
\theoremstyle{plain}
\newtheorem{prop}[thm]{\protect\propositionname}
\theoremstyle{plain}
\newtheorem{lem}[thm]{\protect\lemmaname}

\usepackage[caption=false,font=footnotesize]{subfig}

\@ifundefined{showcaptionsetup}{}{%
 \PassOptionsToPackage{caption=false}{subfig}}
\usepackage{subfig}
\makeatother

\providecommand{\lemmaname}{Lemma}
\providecommand{\propositionname}{Proposition}
\providecommand{\theoremname}{Theorem}

\begin{document}

\title{Profit Maximization Auction and Data Management in Big Data Markets}

\author{Yutao~Jiao, Ping~Wang, Dusit~Niyato, Mohammad~Abu~Alsheikh,
and Shaohan~Feng \\
School of Computer Science and Engineering, Nanyang Technological
University, Singapore 639798}
\maketitle
\begin{abstract}
A big data service is any data-originated resource that is offered over the Internet. The performance of a big data service depends on the data bought from the data collectors. However, the problem of optimal pricing and data allocation in big data services is not well-studied. In this paper, we propose an auction-based big data market model. We first define the data cost and utility based on the impact of data size on the performance of big data analytics, e.g., machine learning algorithms. The big data services are considered as digital goods and uniquely characterized with ``unlimited supply'' compared to conventional goods which are limited. We therefore propose a Bayesian profit maximization auction which is truthful, rational, and computationally efficient. The optimal service price and data size are obtained by solving the profit maximization auction. Finally, experimental results on a real-world taxi trip dataset show that our big data market model and auction mechanism effectively solve the profit maximization problem of the service provider.
\end{abstract}


\section{Introduction}

Over the recent years, big data from various sources, including the Internet of Things (IoT), social network and crowdsouring, have witnessed explosive increase. It is expected that the data value will reach \$92.2 billion by 2026~\cite{http://www.statista.com/}. However, only a small part of today's data is fully utilized and the usage is limited as well. For example, in the petroleum industry, only 1 percent of data from an oil rig with nearly 30,000 sensors is examined~\cite{Manyika2015}. To make profit and increase the data utilization, data can be sold to other organizations. Fortunately, the concepts of data as a service (DaaS) and software as a service (SaaS) have been recently developed. DaaS and SaaS are the core of big data markets where big data and data analytic services are traded and offered over the Internet. The authors in~\cite{NiyatoAlsheikhWangEtAl2016} introduced a typical big data market model composed of three entities, including the data source, service provider, and service customers. The service provider buys the raw data from the data source and applies data analytics on the raw data to create advanced services, e.g., regression and classification models. This paper addresses the following key questions:
\begin{enumerate}
\item How much data should the service provider buy from the data sources?
\item What is the optimal price of a service offered to the customers?
\end{enumerate}
Addressing these questions is important to achieve economic sustainability and maximum profits in data markets.

To answer the aforementioned questions, we propose an auction-based big data market model. Since the big data service is a digital good, we apply the \emph{digital goods auction} for optimal pricing and allocating of digital resources. First, the optimal price and data allocation of the service are obtained by formulating a profit maximization problem as a Bayesian optimal mechanism~\cite{Krishna2009}. Our profit maximization model is truthful, individually rational, and computationally efficient. The optimal data size for maximizing the service's gross profit is derived by solving a convex optimization problem. Second, we analyze the regression problem of taxi trip time prediction to verify the marginal impact of the data size on the customer valuation of the service. Our experimental analysis shows that our auction model is practical and helps the service provider to make purchase and sale strategies. To the best of our knowledge, this is the first paper which applies the digital goods auction in the economics of big data services.

The rest of this paper is organized as follows. Section~\ref{sec:related_work} reviews related work and provides a brief introduction of digital goods auction. The system model of big data market is introduced in Section~\ref{sec:system_model}. Section~\ref{sec:profit_maximization} formulates the profit maximization problem and gives theoretical analysis. Section~\ref{sec:case_study} presents experimental results of the taxi trip time prediction. Finally, Section~\ref{sec:conclusions} concludes the paper.

\section{Related Work\label{sec:related_work}\label{sec:Related-Work}}

Although the study of the economics of big data is still immature, a few papers, e.g.,~\cite{NiyatoHoangLuongEtAl2016,PantelisAija2013,NiyatoLuWangEtAl2016}, addressed the problem of information valuation and data pricing. The author in \cite{Lawrence2012} gives a general formula for defining the value of information which is an important branch in the economic of information. The profit maximization problem has been discussed in many fields, such as cloud computing, smart grid, and cognitive radio networks. A few works have focused on the big data economics involving data collection, processing, and trading. A representative monopolistic business model with two payment methods for IoT information services is proposed in~\cite{GuijarroPlaVidalEtAl2016}. Taking the competition factor into consideration, the authors in~\cite{NiyatoLuWangEtAl2016} present a game theoretical model for substitute and complementary services in the IoT sensing information market. The authors in~\cite{NiyatoAlsheikhWangEtAl2016} develop a subscription-based big data market model. The pricing of bundled services is presented in~\cite{NiyatoHoangLuongEtAl2016}. 

As an effective pricing and resource allocation method in economics, the auction theory has been applied in designing incentive mechanisms to ensure economic sustainability~\cite{YangXueFangEtAl2015,CaoBrahmaVarshney2015}. However, existing pricing approaches based on the conventional auction mechanism are not practical in data services since they are designed for physical goods with limited supply. Digital goods have distinct properties including the unlimited supply and reproduction with almost no marginal cost~\cite{BhattacharjeeGopalMarsdenEtAl2011}. For digital goods, typically the number of items to be sold and the number of customers cannot be determined in advance. The authors in~\cite{PeiKlabjanXie2014} apply a digital goods auction in selling copies of a dataset with the share-averse externality. The authors in~\cite{WangZhengWuEtAl2016} consider the partial competition enabling each bidder to define the list of its competitors. 

Existing works mainly focus on selling the dataset without considering its internality. In this paper, we explore the utility of data, e.g., the data size, and its influence on the service performance and the number of potential buyers. We define the big data service as a digital good. After the successful data collection and analytics, the service provider can sell as many service licenses as there are customers with a neglected marginal cost.

\section{System Model: Big Data and Market Model\label{sec:system_model}}

\begin{figure*}
\begin{centering}
\includegraphics[width=0.8\textwidth]{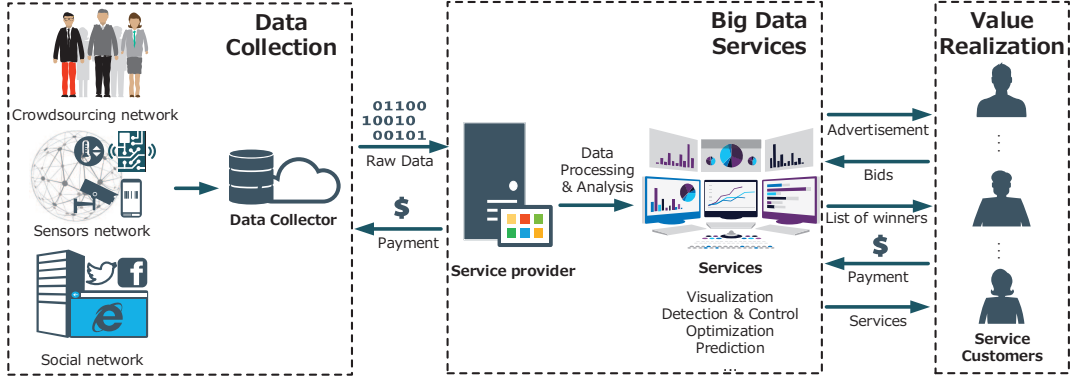}
\par\end{centering}
\caption{Auction-based big data market model \label{fig:AucBGmarket}}
\end{figure*}

\begin{table}[t]
\caption{Frequently used notations.\label{tab:notations}}
\begin{centering}
\begin{tabular}{|c|c|}
\hline 
\textbf{\noun{Notation}} & \textbf{\noun{Description}}\tabularnewline
\hline 
\hline 
$c(q)$ & Cost of $q$ data units\tabularnewline
\hline 
$r(q)$ & Utility of $q$ data units\tabularnewline
\hline 
$k$ & Cost of one data unit\tabularnewline
\hline 
$g$ & Service provider's utility or profit\tabularnewline
\hline 
$p_{i}$ & Payment for customer $i$\tabularnewline
\hline 
$x_{i}$ & Allocation setting of customer $i$\tabularnewline
\hline 
$v_{i}$ & Service valuation of customer $i$\tabularnewline
\hline 
$\phi_{i}(v_{i})$ & Virtual valuation of customer $i$\tabularnewline
\hline 
$M$ & Number of customers\tabularnewline
\hline 
$\Upsilon$ & Influence coefficient\tabularnewline
\hline 
\end{tabular}
\par\end{centering}

\end{table}

Figure~\ref{fig:AucBGmarket} shows the auction-based big data market model considered in this paper. The \emph{data collector} gathers the raw data generated from various sources like sensors and mobile devices. The \emph{service provider} buys the raw data from the data collector and offers big data analytic services over the Internet. The service \emph{customers} are the end users of the data services. Data is collected at different scales and types. Based on the human participation during the data collection, data can be categorized into three classes:
\begin{itemize}
\item Crowdsensing data: People collect data using their personal mobile devices and share the data with the collector. The data collector may pay for the crowdsensing users.
\item Social data: On social networks, people contribute rich data such as text and images.
\item Sensing data: Various sensors, such as GPS, camera and temperature sensors, generate real-time data in sensing systems, e.g., smart transportation.
\end{itemize}
Table \ref{tab:notations} lists frequently used notations used in this paper. We next introduce the data collection cost of big data services. Then, we describe the auction between the service provider and the customers. Then, the utility functions of the service provider and data collectors are provided.

\subsection{Data Collection Cost}

Naturally, the cost of data collection increases substantially as the data size increases. The data collection cost includes energy, time, and hardware resources. It is reasonable to assume that the data cost is monotonically increasing and convex. The data samples are collected into a dataset which contains $N$ data units\footnote{The data unit can be measured in bytes, data sample, or data blocks.}. Thus, the data size which can be bought from the data collector ranges from $0$ to $N$ data units. We introduce a continuous variable $q\in[0,N]$ which denotes the size of raw data sold to the service provider. Thus, we define the data cost function $c(q)$ as follows:

\begin{equation}
c(q)=k\cdot q,\label{eq:CQ}
\end{equation}
where $k>0$ is the cost of one data unit. If the optimal gross profit of the service provider is greater than or equal to $0$, the service provider will buy the data.

\subsection{Big Data Services}

\begin{figure}
\begin{centering}
\includegraphics[width=1\columnwidth]{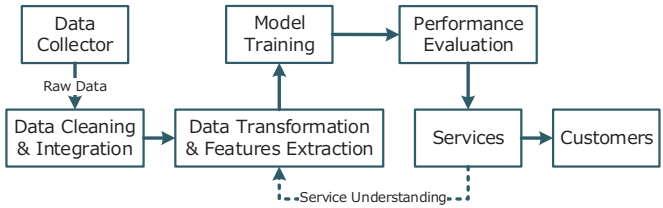}
\par\end{centering}
\caption{Creation of big data services.\label{fig: Bgservice}}
\end{figure}

As shown in Figure~\ref{fig: Bgservice}, there is a common procedure for creating big data services. Data cleaning is first applied to improve the quality of data and remove outlier samples. If the data are collected from multiple sources, removing redundancy in data integration is also necessary. The service provider should transform the data, reduce the dimensions, and extract best features for the model training. 

Classification and regression are two main classes of machine learning schemes. We consider performance measures that are associated with the customer experience. For a classification problem, the\emph{ classification accuracy}, i.e. the proportion of correct prediction results, is used as performance metric. In a regression problem, we incorporate a performance metric called \emph{satisfaction rate} based on the median absolute error~\cite{DraperSmith2014} as follows:

\begin{equation}
r(y,\hat{y})=\frac{n(|y_{i}-\hat{y_{i}}|<\tau)}{L}\label{eq:PefMetric}
\end{equation}
where $\hat{y}_{i}$, $y_{i}$ and $|y_{i}-\hat{y_{i}}|$ are the predicted value, true value, and the absolute prediction error of the $i$-th data sample, receptively. $\tau$ is a preset upper limit constant that represents maximum \emph{tolerance} in prediction quality. The function $n(\cdot)$ counts the number of data samples satisfying the criteria in the bracket. $L$ is the total number of data samples in the dataset. (\ref{eq:PefMetric}) indicates the probability that the prediction error is less than the tolerance level. 

Empirically, we define the service performance metric, e.g., classification accuracy and satisfaction rate, by a data utility function of the data size $q$:

\begin{equation}
r(q;a,b)=a+b\cdot\ln(q)\label{eq:PefQ}
\end{equation}
which is monotone increasing and follows the diminishing marginal utility. $a$ and $b$ are curve fitting parameters of the data utility function $r(q)$ to the real-world experiments. According to~\cite{Domingos2012}, more data can provide better prediction performance. $a$ and $b$ are obtained by nonlinear least squares fitting~\cite{Strutz2010}. Specifically, a series of $N$ experimentation points $(q^{(1)},\alpha^{(1)}),\ldots,(q^{(j)},\alpha^{(j)}),\ldots,(q^{(N)},\alpha^{(N)})$ is performed, where $\alpha^{(j)}$ is the performance metric resulting from a data size of $q^{(j)}$ with $q^{(j+1)}>q^{(j)}$. $r(q;a,b)$ is then found by minimizing the nonlinear least squares as follows:
\[
\min_{a,b}\frac{1}{N}\sum_{1}^{N}||\alpha^{(j)}-r(q^{(j)};a,b)||^{2}.
\]

In Section~\ref{sec:case_study}, we will present a case study of regression-based machine learning algorithms based on real-world datasets to show the validity of the data utility function (\ref{eq:PefQ}). The same analysis can be applied in a classification-based case study, which is omitted from this paper due to space limit. For simplified notations, we use $r(q)$ instead of $r(q;a,b)$ in the rest of the paper. 

\subsection{Value Realization}

Assume there are $M$ customers, where each customer is willing to buy the big data service and has an independent \emph{valuation} of the service. For customer $i$ , the valuation of the service is denoted as $v_{i}$. The service provider first advertises the available service to the customers. Then, the customers submit their sealed bids $\mathbf{b}=(b_{1},\ldots,b_{M})$ which represent their valuations of the offered services $\mathbf{v}=(v_{1},\ldots,v_{M})$. After receiving the bids, the service provider determines the list of winners containing the allocation $\mathbf{x}=(x_{1},\ldots,x_{M})$ and prices $\mathbf{p}=(p_{1},\ldots,p_{M})$. The setting $x_{i}=1$ indicates customer $i$ is within the winner list and being allocated service while $x_{i}=0$ is for no service. $p_{i}$ is the sale price that customer $i$ is charged by the service provider. At the end of the auction, the winners make the payment and access the big data service. 

With the aforementioned setting, the utility of the service provider $g(\mathbf{x},\mathbf{p})$ and the $i$-th customer $u_{i}$ are expressed as follows:
\begin{equation}
g(\mathbf{x},\mathbf{p})=\sum_{i=1}^{M}x_{i}\cdot p_{i}-c(q),\label{eq:uDP}
\end{equation}
\begin{equation}
u_{i}=v_{i}\cdot x_{i}-p_{i}.\label{eq:Uci}
\end{equation}
As defined in (\ref{eq:uDP}), the service provider sets the sale price and the data size to maximize its profit. The natural objective of the customer is to choose a bid that maximizes its utility defined in (\ref{eq:Uci}) as the difference between its valuation and price.

\section{Bayesian Profit Maximization Auction\label{sec:profit_maximization}}

The calculation of the winner list is critical for a successful and efficient big data service trade. In this section, we focus on maximizing the service provider's gross profit. We first apply a Bayesian digital goods auction to calculate the optimal sale price of the service when the data size is fixed. Then we derive the optimal solution of the requested data size by solving a convex optimization problem.

\subsection{Valuation Distribution}

Here, we first define customer $i$'s service valuation $v_{i}$ in the big data market as follows:
\[
v_{i}=d_{i}\cdot r(q)\cdot\gamma
\]
where $d_{i}\in[0,1]$ is the degree of that particular customer personal\emph{ }preference. High preference degree indicates high dependence or demand on the data service. $d_{i}$ is defined based on the customer needs, habit, and salary. For example, a frequent traveler has a high degree of preference for weather forecast services compared to an office employee. $\gamma\in(0,\infty)$ is the influence coefficient defining the impact of the service performance on the customer valuation. The final valuation, i.e., submitted bid, is jointly determined by the degree of preference and service performance. For simplicity, we assume that $d_{i}$ is a random variable sampled independently from the uniform distribution with a range of $[0,1]$. Then, the probability density function $f(v)$ and cumulative distribution function $F(v)$ of the the customer valuation can be derived as follows:
\[
f(v)=\begin{cases}
\frac{1}{r(q)\cdot\gamma} & v\in[0,r(q)\cdot\gamma],\\
0 & \text{otherwise}.
\end{cases}
\]
\begin{equation}
F(v)=P(V\leqslant v)=\begin{cases}
0 & v\in(-\infty,0),\\
\frac{v}{r(q)\cdot\gamma} & v\in[0,r(q)\cdot\gamma],\\
1 & v\in(r(q)\cdot\gamma,\infty).
\end{cases}\label{eq:CDF}
\end{equation}

\subsection{Optimal Sale Price\label{subsec:Optimal-Sale-Price}}

In our Bayesian formulation, the customer valuation $\mathbf{v}$ are drawn independently from the distribution $F(v)$ given in (\ref{eq:CDF}). We define the \emph{virtual valuation} of customer $i$ as $\phi_{i}(v_{i})=v_{i}-\frac{1-F(v_{i})}{f(v_{i})}$. The \emph{virtual surplus} of the service provider\emph{ }can be denoted as $\sum_{i=1}^{M}x_{i}\cdot\phi_{i}(v_{i})-c(q)$. From~(\ref{eq:CDF}), we can note that the hazard rate of the distribution, i.e., $\frac{f(\cdot)}{1-F(\cdot)}$, is monotone non-deceasing which implies the virtual valuations are monotone non-decreasing as well. This satisfies the necessary and sufficient condition for the truthfulness of the virtual surplus maximization~\cite{NisanRoughgardenTardosEtAl2007}.

We next introduce the profit maximization problem based on the Myerson's optimal mechanism~\cite{Myerson1981}. This enables deriving the expected gross profit as a virtual surplus maximization problem.
\begin{prop}
\label{thm:The-expected-profit}The expected profit of any truthful
mechanism $(\mathbf{p},\mathbf{x})$ is equal to its expected virtual
surplus, i.e., $\mathbf{E}_{\mathbf{v}}[g(\mathbf{x}(\mathbf{v}),\mathbf{p}(\mathbf{v}))]=\mathbf{E}_{\mathbf{v}}[\sum_{i=1}^{M}x_{i}(\mathbf{v})\cdot\phi_{i}(v_{i})-c(q)]$.\end{prop}
\begin{IEEEproof}
This result follows from the Myerson's lemma.\end{IEEEproof}
\begin{lem}
(Myerson's Lemma) For any truthful mechanism $(\mathbf{p},\mathbf{x})$, the expected payment of bidder $i$ with valuation distribution $F$ satisfies:
\[
\mathbf{E}_{b_{i}}[p_{i}(b_{i})]=\mathbf{E}_{b_{i}}[x_{i}\cdot\phi_{i}(b_{i})]
\]
where $b_{i}=v_{i}.$
\end{lem}

The optimal mechanism is described as follows:
\begin{enumerate}
\item Receive the sealed bids $\mathbf{b}$ and compute the customer's \emph{virtual bids}: $b'_{i}=\phi_{i}(b_{i})=b_{i}-\frac{1-F(b_{i})}{f(b_{i})}$. 
\item Apply the Vickrey\textendash Clarke\textendash Groves (VCG) auction~\cite{Krishna2009} on virtual bids $b'$ and output the allocation $\mathbf{x}'$ and the virtual payment $\mathbf{p}'$ which maximize the virtual surplus.
In this step, the virtual payment is computed by 
\begin{figure}[H]
$p_{i}'=\begin{cases}
0 & x'_{i}=0,\\
\min\{\sum_{j\in W(\mathbf{b}_{-i}),j\neq i}\phi_{j}-\sum_{j\in W(\mathbf{b}),j\neq i}\phi_{j},0\} & x'_{i}=1,
\end{cases}$
\end{figure}
where $W(\mathbf{b})$ is the set of winners that are allocated items and $W(\mathbf{b}_{-i})$ is the set calculated by the VCG mechanism among all except the customer $i$.
\item Calculate the final allocation $\mathbf{x}=\mathbf{x'}$ and payment $\mathbf{p}$ with $p_{i}=\phi_{i}^{-1}(p'_{i})$.
\end{enumerate}
Since big data services are digital goods that have unlimited supply and almost no marginal cost, we can allocate the service to customer $i$ as long as $b'_{i}\geqslant0$ in the step $2$. Here, the actual payment that the winning customer must take is the minimum bid, i.e., $\inf\{b:\phi(b_{i})\geqslant0\}$, which is the solution for $\phi(b)=b-\frac{1-F(b)}{f(b)}=0$. Hence, according to Theorem \ref{thm:The-expected-profit}, the service provider can offer customers this optimal sale price, denoted by $p^{*}=\phi^{-1}(0)$, to maximize his profit in expectation. The Bayesian digital goods auction has three desirable properties: 
\begin{itemize}
\item Incentive compatibility: Since the payment required for customer $i$ solely depends on other customers' bids in the VCG auction, the auction mechanism guarantees that every customer can achieve the best outcome just by bidding its true valuation, i.e., $b_{i}=v_{i}$. Being truthful can curb the market speculation and reduce the unnecessary cost on making bidding strategy.
\item Individual rationality: Each customer will have a non-negative utility by submitting its true valuation. 
\item Computational efficiency: The list of winners can be computed in polynomial time, which has the complexity only of $O(1)$ per customer.
\end{itemize}

\subsection{Optimal Data Size}

Since the proposed auction mechanism is truthful, the customer $i$'s bid is equal to its valuation, i.e., $b_{i}=v_{i}$. Based on the optimal mechanism in \ref{subsec:Optimal-Sale-Price}, we can obtain the optimal sale price with predefined valuation distribution $F(v)$:

\begin{equation}
p^{*}=\phi^{-1}(0)=\frac{\gamma\cdot r(q)}{2}.\label{eq:popt}
\end{equation}

Then, an optimization problem can be formulated in order to obtain the optimal size of raw data which are bought from the data collector. Substituting $c(q)$ from (\ref{eq:CQ}), $r(q)$ from (\ref{eq:PefQ}) and $p_{i}=p^{*}$ from (\ref{eq:popt}) into (\ref{eq:uDP}), the expected utility of the service provider is computed as follows:

\begin{align}
\mathbf{E_{v}}[g(q)] & =\begin{cases}
0 & q=0,\\
M\cdot P(V\geqslant p^{*})\cdot p^{*}-k\cdot q & q>0.
\end{cases}\nonumber \\
 & =\begin{cases}
0 & q=0,\\
\frac{M\cdot\gamma\cdot(a+b\cdot\ln(q))}{4}-k\cdot q & q>0.
\end{cases}\label{eq:UdpQplugged}
\end{align}

\begin{prop}
There exists a globally optimal data size $q^{*}$ that maximizes the service provider's expected utility in (\ref{eq:UdpQplugged}) over $q\in[0,N]$.
\end{prop}
\begin{IEEEproof}
When the utility of the service provider is positive $g(q)>0$, we can find the second derivative of $g(q)$ as follows:
\begin{equation}
\frac{\partial^{2}g\left(\cdot\right)}{\partial q^{2}}=-\frac{M\cdot\gamma\cdot b}{4\cdot q^{2}}.\label{eq:2ndderivative}
\end{equation}
Since $q>0$ and $a,b,\gamma,M>0$, it can be shown that (\ref{eq:2ndderivative}) is always non-positive. Thus, the utility function $g$ is a concave function for $q\in(0,N]$. By differentiating $g(q)$ with respect to $q$, we have 

\[
\frac{\partial g\left(\cdot\right)}{\partial q}=\frac{M\cdot\gamma\cdot b}{4\cdot q}-k.
\]
The optimal solution $q_{+}^{*}$ can be efficiently defined by solving $\frac{dg}{dq}=0$. We can get the closed-form solution of $q_{+}^{*}$ as follows:
\begin{equation}
q_{+}^{*}=\begin{cases}
\frac{M\cdot\gamma\cdot b}{4\cdot k} & 0<\frac{M\cdot\gamma\cdot b}{4\cdot k}<N,\\
N & \frac{M\cdot\gamma\cdot b}{4\cdot k}\geq N.
\end{cases}
\end{equation}
When the utility of the service provider is non-positive $g(q)\leqslant0$, the service provider will reject to buy the data. Accordingly, there exists a globally optimal data size $q^{*}\in[0,N]$ which can be expressed as follows:
\[
q^{*}=\begin{cases}
0, & g(q)\leqslant0,\\
q_{+}^{*}, & g(q)>0.
\end{cases}
\]
From these results, we can find that the service provider can reject to buy the data, i.e, $q^{*}=0$, if the data cost is too high.
\end{IEEEproof}

\section{Experimental Results: Taxi Trip Time Prediction\label{sec:case_study}}

In this section, we provide a case study along with representative numerical results of the proposed auction, from which we can further obtain useful decision making strategies for the service provider.

\subsection{Experiment Setup }

We use a real-world taxi service trajectory dataset \cite{Moreira-MatiasGamaFerreiraEtAl2013} to develop a data service that predicts the trip time for each taxi driver. The taxi service trajectory dataset includes $442$ drivers and $L=1,710,671$ taxi trip samples. Each sample contains taxi geolocation data collected by a vehicular GPS and relevant information, such as trip ID, taxi ID, and time-stamp. We first pre-process the raw data by removing the fault data samples and extract valuable features as well as corresponding labels. Totally, we prepare $1,160,815$ samples for model training and $501,858$ samples for testing and performance evaluation. In our experiment, we use a classical machine learning algorithm, i.e., random forest regression, for data analytics. We assume a base of $M=10000$ customers. For demonstration purposes, we normalize the data size $q$ from $0-N$ range to $0-100$ range in this section.

\subsection{Verification for Data Utility Function}

\begin{figure}
\centering{}\includegraphics[width=1\columnwidth,height=0.33\textheight]{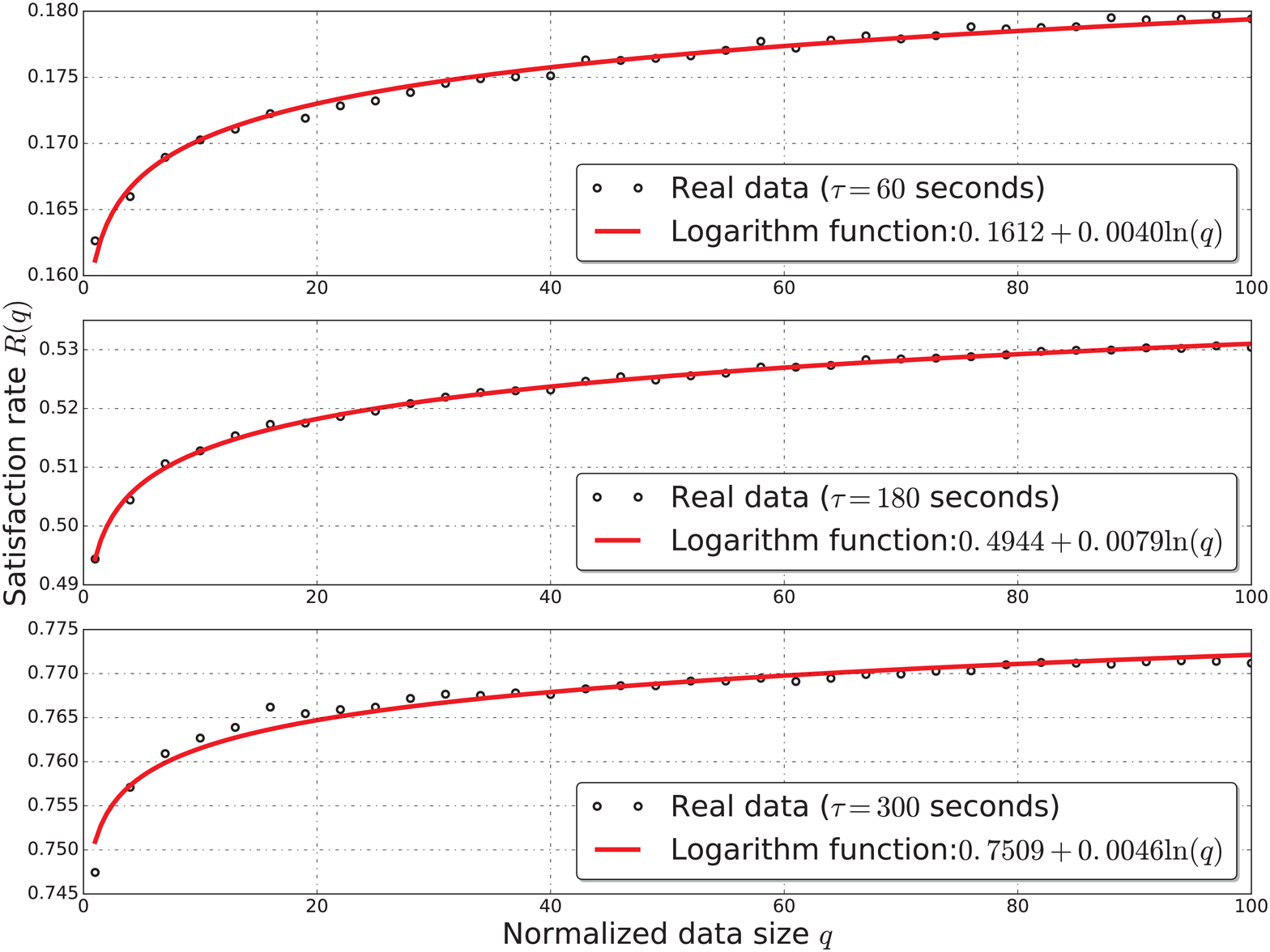}\caption{Estimation of the data utility function $r(q)$ using random forest
regression under different tolerance values. (Top: $60$ seconds.
Middle: $180$ seconds. Bottom: $300$ seconds.)\label{fig:DataUtility}}
\end{figure}

\begin{figure}[h]
\subfloat[\label{fig:p2uDP}]{\includegraphics[width=0.49\columnwidth]{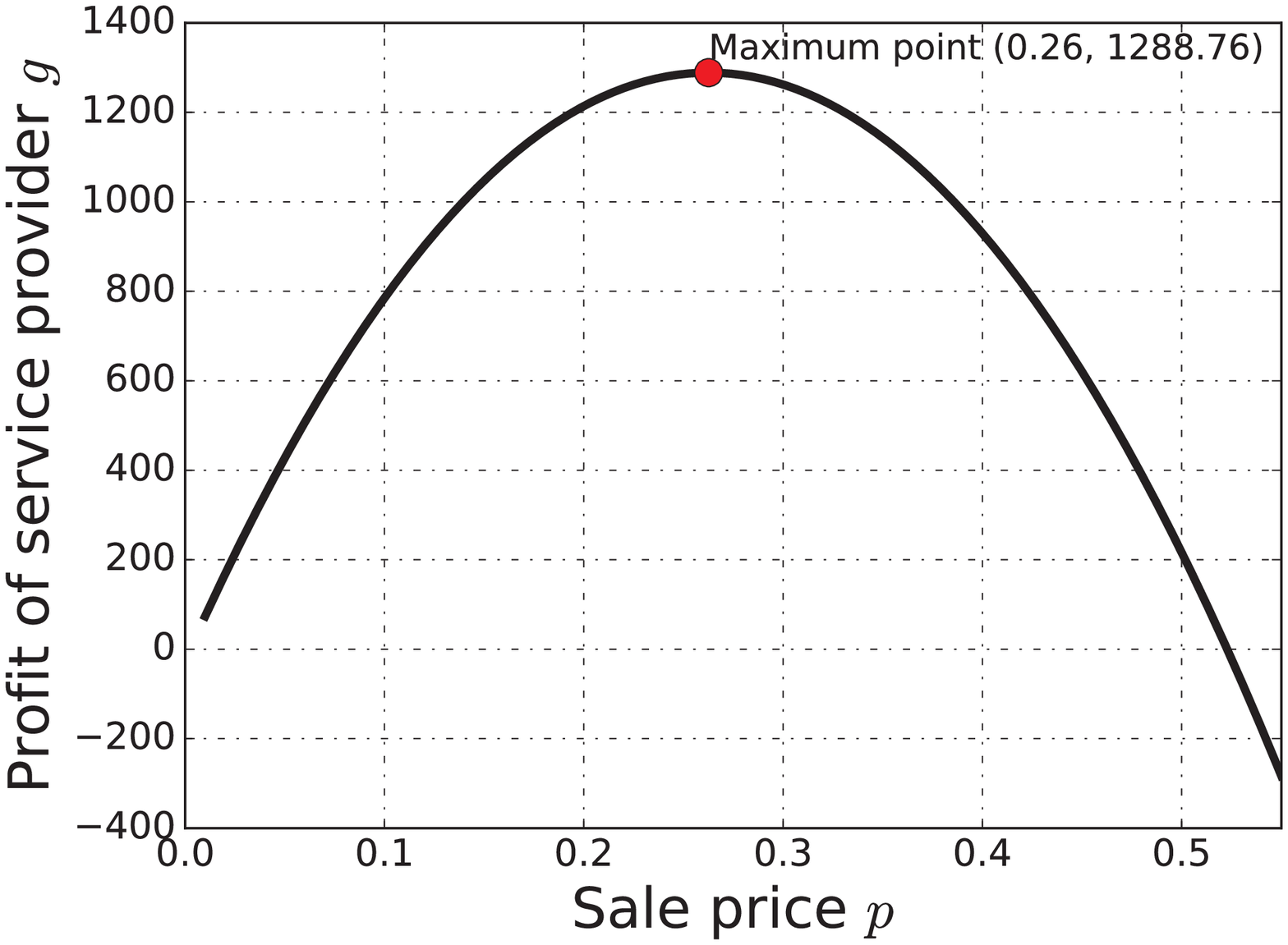}

}\subfloat[\label{fig:Q2udp}]{\includegraphics[width=0.5\columnwidth]{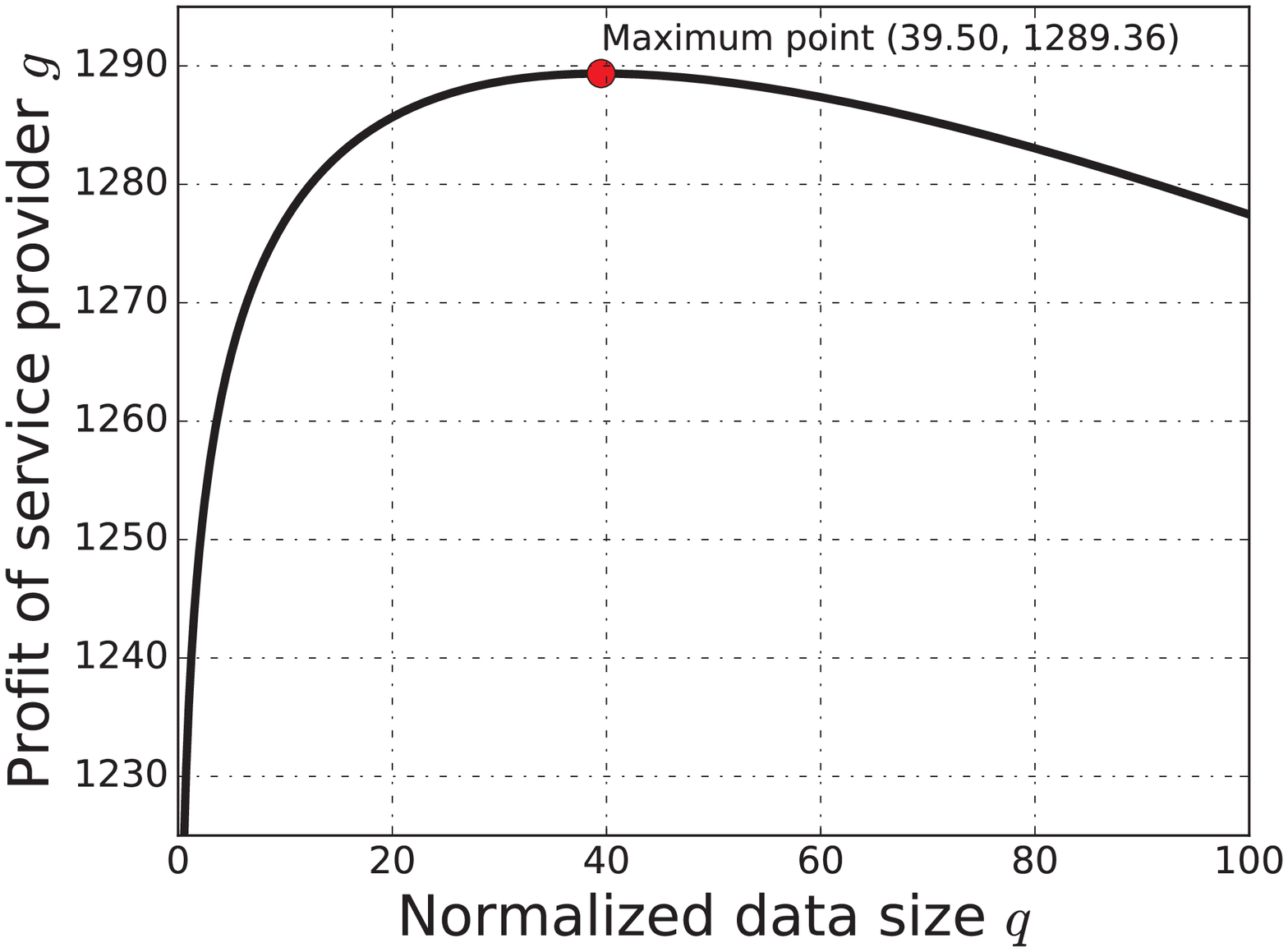}

}
\caption{Profit of service provider $g$. (a) Impact of $p$ on $g$. (b) Impact
of $q$ on $g$.\label{fig:4}}
\end{figure}

\begin{figure}[h]
\subfloat[\label{fig:k2udp}]{\includegraphics[width=0.49\columnwidth]{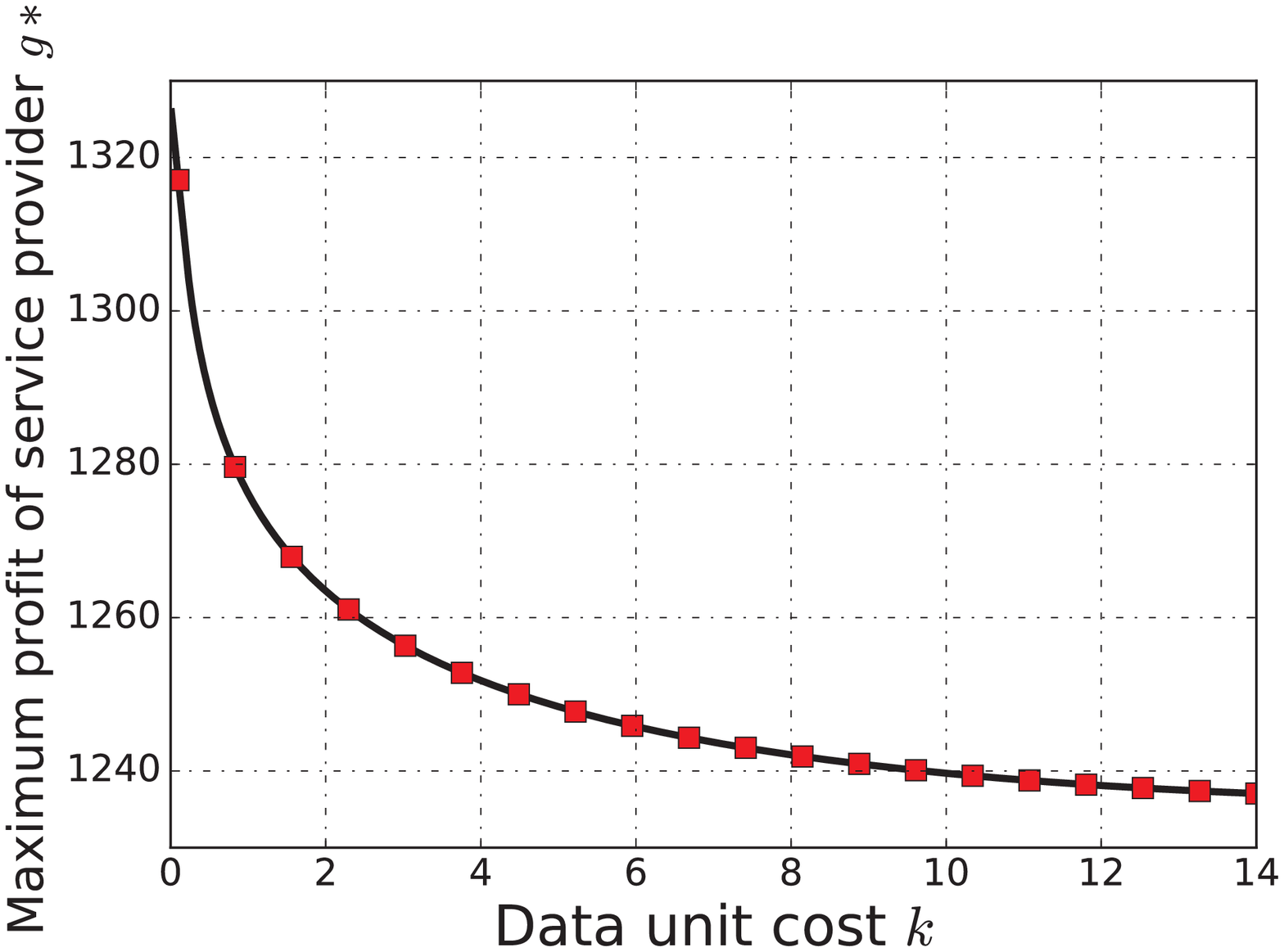}

}\subfloat[\label{fig:I2udp}]{\includegraphics[width=0.5\columnwidth]{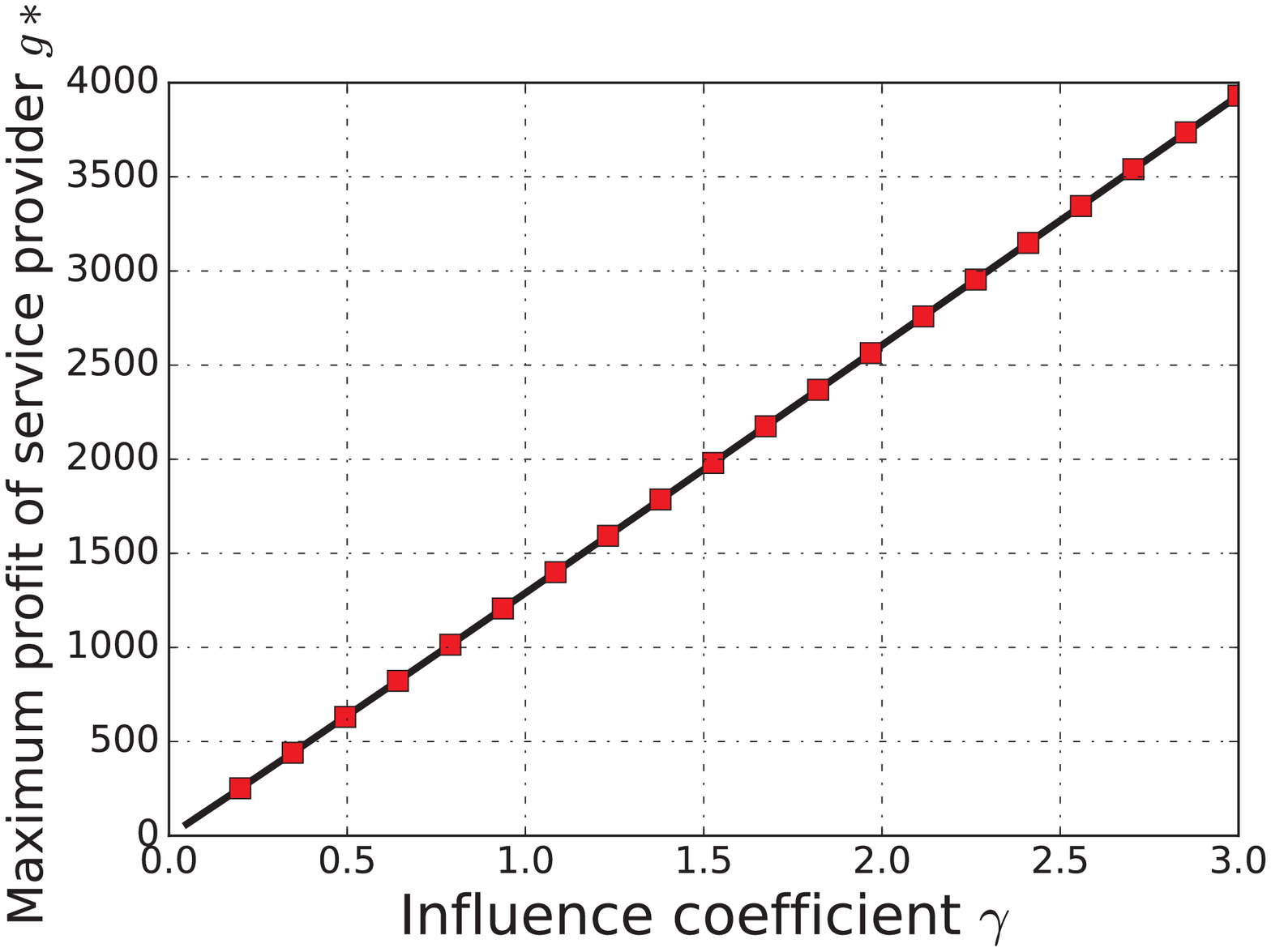}

}

\caption{Maximum profit of service provider $g^{*}$ under varied (a)~data
unit cost $k$ and (b)~influence coefficient $\gamma$. \label{fig:5}}
\end{figure}

We use the performance metric \emph{satisfaction rate} defined in (\ref{eq:PefMetric}) to evaluate our data service. For each taxi driver, the less the difference between the predicted result and true trip time, the faster the driver will pick up another passenger, which increases its revenues. We respectively set $\tau=60$, $180$, or $300$, where $60$ seconds ($1$ minute), $180$ seconds ($3$ minutes), $300$ seconds ($5$ minutes) are the common tolerance values for a person to wait for a taxi service. Figure \ref{fig:DataUtility} shows the change of the service performance under different amount of requested data. The service performance increases as the data size increases, but meanwhile the increase of the service performance becomes diminishing. More importantly, we note that the performance utility function defined in (\ref{eq:PefQ}) can well fit the actual performance results which demonstrate the diminishing returns. From these results, we choose the tolerance of $180$ seconds and use $r(q)=0.4944+0.0079\ln(q)$ in the rest of this section.

\begin{figure}
\subfloat[\label{fig:k2Q}]{\includegraphics[width=0.5\columnwidth]{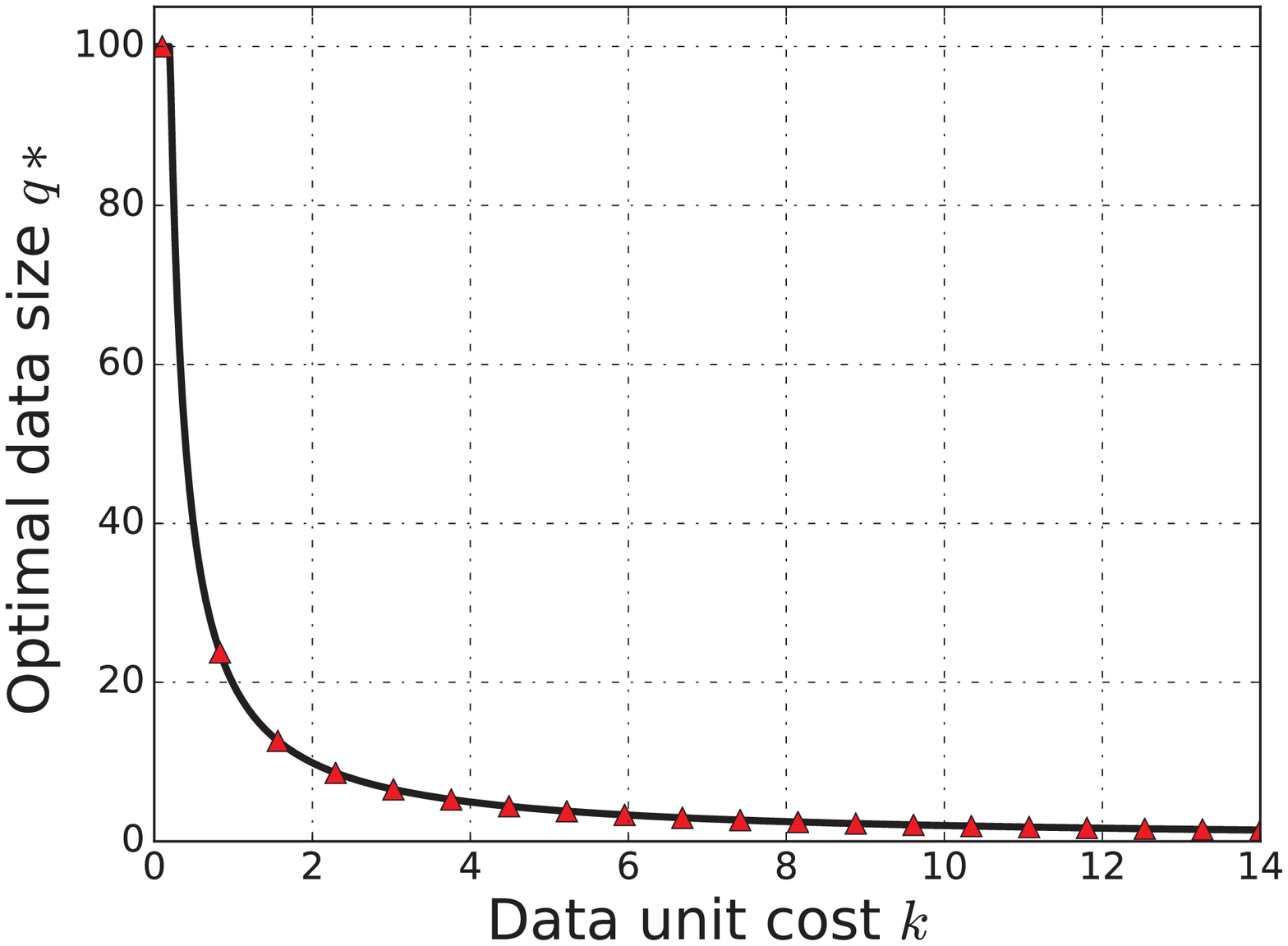}

}\subfloat[\label{fig:I2Q}]{\includegraphics[width=0.5\columnwidth]{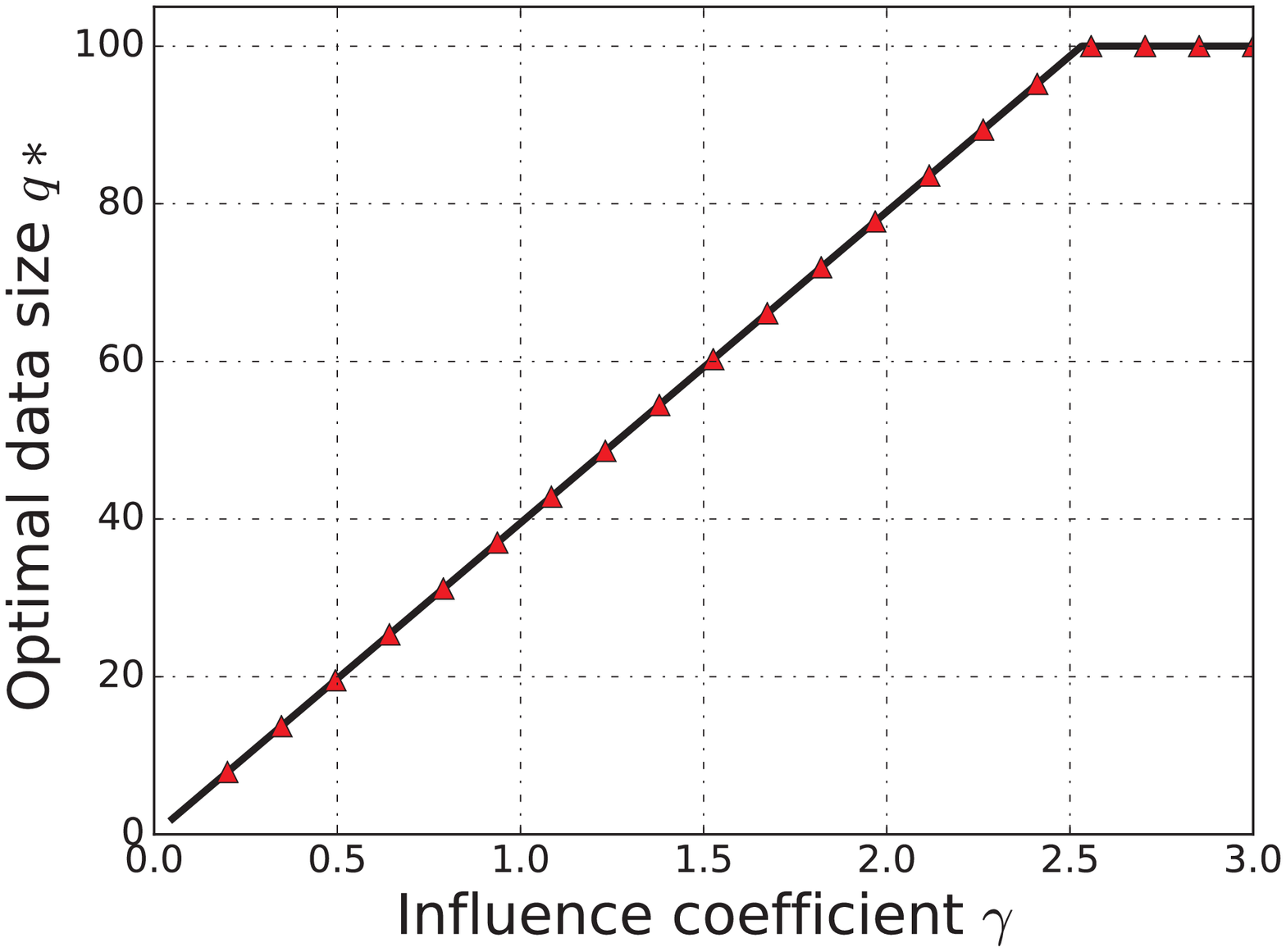}

}

\caption{Optimal requested data size $q^{*}$ under varied (a)~data unit cost
$k$ and (b)~influence coefficient $\gamma$.\label{fig:k2QandI2Q}}
\end{figure}

\subsection{Numerical Results and Strategies for Decision Making}
\begin{enumerate}
\item Expected Profit of the Service Provider: Figure~\ref{fig:4} and Figure~\ref{fig:5} show the impacts of $p$, $q$, $k$ and $\gamma$ on the service provider's profit. In Figure~\ref{fig:p2uDP}, we fix $q=50$, $k=0.5$ and $\gamma=1$ while varying the value of sale price $p$. Apparently, the optimal sale price that maximizes the profit is exactly equal to the value calculated by (\ref{eq:popt}). In Figure~\ref{fig:Q2udp}, we fix $k=0.5$ and $\gamma=1$. When the data size is small, the service performance is poor and the optimal sale price is low. Thus, the service provider's profit is small. However, if the data size is large, the service provider has to pay more for the raw data which causes the decrease of its profit. Clearly, there is an optimal profit $g^{*}$ that can be achieved when the optimal requested data size is applied. In Figure~\ref{fig:k2udp}, we fix $\gamma=1$. The optimal service provider's profit $g^{*}$ deceases as the unit cost of data $k$ increases and tends to be zero when $k$ is too high. In Figure~\ref{fig:I2udp}, we fix $k=0.5$. We observe that the optimal service provider's profit $g^{*}$ increases linearly as the influence coefficient $\gamma$ increases. The more impact of service performance on the customer valuation, the more profit the service provider can achieve.
\item Optimal Data Size $q^{*}$: Figure~\ref{fig:k2QandI2Q} shows the impact of $k$ and $\gamma$ on the optimal requested data size. In Figure~\ref{fig:k2Q}, as the unit cost of data increases, the optimal data size bought from data collector decreases. If the data unit cost $k$ is quite low, the service provider should always buy all the collector's data. However, if data unit cost $k$ is too high and the service provider will suffer deficit, the best strategy for service provider is not to buy the data. In Figure~\ref{fig:I2Q}, we observe that the value of $q^{*}$ increases linearly with the influence coefficient $\gamma$. When $\gamma$ becomes large enough, the $q^{*}$ reaches and remains the maximum size $100$. This can be interpreted as when the impact of service performance gets significant, the service provider should buy as much data as it could. 
\end{enumerate}

\section{Conclusions\label{sec:conclusions}}

In this paper, we have developed an auction-based big data market model. In the model, we have proposed a Bayesian digital goods auction to allocate services to customers and define the optimal sale price for each winner. We have proved that the mechanism is truthful, individually rational and computationally efficient. Our optimization model maximizes the service provider's profit by choosing the optimal sale price and data size bought from the data collector. Based on a real-world dataset, we have verified our profit maximization auction and provided numerical results by a case study on taxi trip time prediction.

\section*{Acknowledgment}
This work was supported by Singapore MOE Academic Research Fund Tier 2 (MOE2013-T2-2-070).

\bibliographystyle{IEEEtran}

\end{document}